\documentclass[11pt]{article}
\usepackage[dvips]{color}
\usepackage{epsfig}
\usepackage{amsmath}
\usepackage{graphicx}
\date{}
\begin{document}
\title{{\bf Classical polymerization of the Schwarzschild metric}}
\author{Babak Vakili\thanks{email: b.vakili@iauctb.ac.ir}\\\\{\small {\it  Research
Institute for Astronomy and Astrophysics of Maragha (RIAAM)-
Maragha, IRAN, P. O. Box: 55134-441}}\\ and
\\{\small {\it Department of Physics, Central
Tehran Branch, Islamic Azad University, Tehran, Iran}}} \maketitle

\begin{abstract}
We study a spherically symmetric setup consisting of a Schwarzschild
metric as the background geometry in the framework of classical
polymerization. This process is an extension of the polymeric
representation of quantum mechanics in such a way that a
transformation maps classical variables to their polymeric
counterpart. We show that the usual Schwarzschild metric can be
extracted from a Hamiltonian function which in turn, gets
modifications due to the classical polymerization. Then, the polymer
corrected Schwarzschild metric may be obtained by solving the
polymer-Hamiltonian equations of motion. It is shown that while the
conventional Schwarzschild space-time is a vacuum solution of the
Einstein equations, its polymer-corrected version corresponds to an
energy-momentum tensor that exhibits the features of dark energy. We
also use the resulting metric to investigate some thermodynamical
quantities associated to the Schwarzschild black hole, and in
comparison with the standard Schwarzschild metric the similarities
and differences are discussed.
\vspace{5mm}\noindent \vspace{0.8mm}\newline
\end{abstract}
\section{Introduction}
One of the most important arenas that shows the power of general
relativity in describing the gravitational phenomena is the
classical theory of black hole physics. However, when we introduce
the quantum considerations to study of a gravitational systems,
general relativity does not provide a satisfactory description of
the physics of the system. The Phenomena such as black hole
radiation and all kinds of cosmological singularities are among the
phenomena in which the use of quantum mechanics in their description
is inevitable. This means that although general relativity is a
classical theory, in its most important applications, the system
under consideration originally obeys the rules of quantum mechanics.
Therefore, any hope in the accurate description of gravitational
systems in high energies depends on the development of a complete
theory of quantum gravity. That is why the quantum gravity is one of
the most important challenges in theoretical physics which from its
DeWitt's traditional canonical formulation \cite{De Witt} to the
more modern viewpoints of string theory and loop quantum gravity
(LQG) \cite{Loop}-\cite{Rov} has gone a long way. One of the main
features of the space-time proposed in LQG is its granular structure
which in turn, supports the idea of existence of a minimal
measurable length. In the absence of a full theory of quantum
gravity, effective theories which somehow exhibits quantum effects
in gravitational systems play a significant role. These are theories
which show some phenomenological aspects of quantum gravity and
usually use a certain deformation in their formalism. For example,
theories like generalized uncertainty principles and noncommutative
geometry are is this category \cite{GUP1}-\cite{NCG1}.

Among the effective theories that also use a minimal length scale in
their formalism, we can mention the polymer quantization
\cite{Corichi1}, which uses the methods very similar to the
effective theories of LQG \cite{Ashtekar}. In polymer quantum
approach a polymer length scale, $\lambda$, which shows the scale of
the segments of the granular space enters into the Hamiltonian of
the system to deform its functional form into a so-called polymeric
Hamiltonian. This means that in a polymeric quantized system in
addition of a quantum parameter $\hbar$, which is responsible to
canonical quantization of the system, there is also another quantum
parameter $\lambda$, that labels the granular properties of the
underlying space. This approach then opened a new windows for the
theories which are dealing with the quantum gravitational effects in
physical systems such as quantum cosmology and black hole physics,
see for instance \cite{qc pol} and \cite{pol bl} and the references
therein.

To polymerize a dynamical system one usually begins with a classical
system described by Hamiltonian $H$. The canonical quantization of
such a system transforms its Hamiltonian to an Hermitian operator,
which now contains the parameter $\hbar$, in such a way that in the
limit $\hbar \rightarrow 0$, the quantum Hamiltonian $H_{\hbar}$
returns to its classical counterpart. By polymerization, the
Hamiltonian gets an additional quantum parameter $\lambda$, which is
rooted in the ideas of granular structure of the space-time.
Therefore, by taking the classical limit of the resulting
Hamiltonian $H_{\hbar,\lambda}$, we arrive at a semi-classical
theory in which the parameter $\lambda$ is still present. To achieve
the initial classical theory, one should once again take the limit
$\lambda \rightarrow 0$ from this intermediate theory. It is
believed that such an effective classical theories $H_{\lambda}$,
have enough rich structure to exhibit some important features of the
system related to the quantum effects without quantization of the
system. The process by which the theory $H_{\lambda}$ is obtained
from the classical theory is called {\it classical polymerization}.
A detailed explanation of this process with some of its cosmological
applications can be found in \cite{CPR}.

In this paper, we are going to study how the metric of the
Schwarzschild black hole gets modifications due to the classical
polymerization. Since the thermodynamical properties of the black
hole come from its geometrical structure, the corrections to the
black hole's geometry yield naturally modifications to its
thermodynamics. To do this, we begin with a general form of a
spherically symmetric space-time and then construct a Hamiltonian in
such a way that the Schwarzschild metric is resulted from the
corresponding Hamiltonian equations of motion. We then follow the
the procedure described above and by application it to the mentioned
Hamiltonian we get the classical polymerized Hamiltonian, by means
of which we expect to obtain the polymer-corrected Schwarzschild
metric. The paper is organized as follows. In section 2 we have
presented a brief review of the polymer representation and classical
polymerization. Section 3 is devoted to the Hamiltonian formalism of
a general spherically symmetric space time. We show in this section
that the resulting Hamiltonian equations of motion yield the
Schwarzschild solution. In section 4, we will apply the classical
polymerization on the Hamiltonian of the spherically symmetric
space-time given in section 3 to get the polymerized Hamiltonian. We
then construct the deformed Hamiltonian equations of motion and
solve them to arrive the polymer corrected Schwarzschild metric. The
energy-momentum tensor of the matter field corresponding to this
metric as well as some of its thermodynamical properties are also
presented in this section. The radial geodesics of the light and
particles are obtained in section 5 and finally, we summarize the
results in section 6.

\section{Classical polymerization: a brief review}
As is well-known, in Schr\"{o}dinger picture of quantum mechanics,
the coordinates and momentum representations are equivalent and may
be easily converted to each other by a Fourier transformation.
However, in the presence of the quantum gravitational effects the
space-time may take a discrete structure so that such a well-defined
representations are no longer applicable. As an alternative, polymer
quantization provides a suitable framework for studying these
situations \cite{Corichi1,Ashtekar}. The Hilbert space of this
representation of quantum mechanics is ${\mathcal H}_{\rm
poly}=L^2(R_{_d},d\mu_{_d})$, where $d\mu_{_d}$ is the Haar measure,
and $R_{_d}$ denotes the real discrete line whose segments are
labeled by an extra dimension-full parameter $\lambda$ such that the
standard Schr\"{o}dinger picture will be recovered in the continuum
limit $\lambda\rightarrow\ 0$. This means that by a classical limit
$\hbar\rightarrow\,0$, the polymer quantum mechanics tends to an
effective $\lambda$-dependent classical theory which is somehow
different from the classical theory from which we have started. Such
an effective theory may also be obtained directly from the standard
classical theory, without referring to the polymer quantization, by
using of the Weyl operator \cite{CPR}. The process is known as {\it
polymerization} with which we will deal in the rest of this paper.

According to the mentioned above form of the Hilbert space of the
polymer representation of quantum mechanics, the position space
(with coordinate $q$) has a discrete structure with discreteness
parameter $\lambda$. Therefore, the associated momentum operator
$\hat{p}$, which is the generator of the displacement, does not
exist \cite{Ashtekar}. However, the Weyl exponential operator (shift
operator) correspond to the discrete translation along $q$ is well
defined and effectively plays the role of momentum associated to $q$
\cite{Corichi1}. This allows us to utilize the Weyl operator to find
an effective momentum in the semiclassical regime. So, consider a
state $f(q)$, its derivative with respect to the discrete position
$q$ may be approximated by means of the Weyl operator as \cite{CPR}

\begin{eqnarray}\label{FWD}
\partial_{q}f(q)\approx\frac{1}{2\lambda}[f(q+\lambda)-f(q-
\lambda)]\hspace{2cm}\nonumber\\=\frac{1}{2\lambda}\Big(
\widehat{e^{ip\lambda}}-\widehat{e^{-ip\lambda}}\Big)\,f(q)=
\frac{i}{\lambda}\widehat{\sin(\lambda p)}\,f(q),
\end{eqnarray}
and similarly the second derivative approximation will be
\begin{eqnarray}\label{SWD}
\partial_{q}^2f(q)\approx\frac{1}{\lambda^2}[f(q+\lambda)-2
f(q)+f(q-\lambda)]\hspace{1cm}\nonumber\\=\frac{2}{\lambda^2}
(\widehat{\cos(\lambda p)}-1)\,f(q).\hspace{2cm}
\end{eqnarray}
Having the above approximations at hand, we define the
polymerization process for the finite values of the parameter
$\lambda$ as

\begin{eqnarray}\label{Polymerization}
\hat{p}\rightarrow\,\frac{1}{\lambda}\widehat{\sin(\lambda p)},
\hspace{1cm}\hat{p}^2\rightarrow\,\frac{2}{\lambda^2}(1-
\widehat{\cos(\lambda p)}).
\end{eqnarray}This replacements suggest the idea that a
classical theory may be obtained via this process, but now without
any attribution to the Weyl operator. This is what which is dubbed
usually as {\it classical Polymerization} in literature
\cite{Corichi1,CPR}:

\begin{eqnarray}\label{PT}
q\rightarrow q,\hspace{1.5cm}p\rightarrow\frac{ \sin(\lambda
p)}{\lambda},\hspace{5mm}p^2\rightarrow
\frac{2}{\lambda^2}\left[1-\cos(\lambda p)\right],
\end{eqnarray}where now $(q,p)$ are a pair of classical phase space
variables. Hence, by applying the transformation (\ref{PT}) to the
Hamiltonian of a classical system we get its classical polymerized
counterpart. A glance at (\ref{PT}) shows that the momentum is
periodic and varies in a bounded interval as
$p\in[-\frac{\pi}{\lambda},+\frac{\pi}{ \lambda})$. In the limit
$\lambda\rightarrow\,0$, one recovers the usual range for the
canonical momentum $p\in(-\infty,+\infty)$. Therefore, the
polymerized momentum is compactified and topology of the momentum
sector of the phase space is $S^1$ rather than the usual $R$
\cite{NaturalCutoff}. Our set-up to explain the classical
polymerization of a dynamical system is now complete. In section 4,
we will return to this issue by some more explanations and apply it
to the Hamiltonian dynamics of a spherically symmetric space-time.

\section{Hamiltonian model of the spherically symmetric space-time}
We start with the general spherically symmetric line element as
\footnote{It can be shown that by introducing of new radial and time
coordinates as $b(r)\rightarrow r'$, and
$I(r)[a(r)dt-B(r)dr]\rightarrow dt'$, this metric takes the standard
form of static spherically symmetric line elements:
$ds^2=-A(r)dt^2+C(r)dr^2+r^2(d\vartheta^2+\sin^2 \vartheta
d\varphi^2)$.} \cite{3}

\begin{equation}\label{A}
ds^2=-a(r)dt^2+N(r)dr^2+2B(r)dtdr+b^2(r)\left(d\vartheta^2+\sin^2\vartheta
d\varphi^2\right),
\end{equation}where $a(r)$, $B(r)$, $N(r)$
and $b(r)$ are some functions of $r$. Upon substitution this metric
into the Einstein-Hilbert action

\begin{eqnarray}\label{EH}
{\cal S}=\frac{1}{16\pi G}\int d^4x \sqrt{-g}{\cal R},
\end{eqnarray}the action takes the form

\begin{equation}\label{EH1}
{\cal S}=\int dt \int dr L(a,b,n),\end{equation}where \cite{3}

\begin{equation}\label{C}
L=2\sqrt{n}\left(\frac{a'b'b}{n}+\frac{ab'^2}{n}+1\right),
\end{equation}
is an effective Lagrangian in which the primes denote
differentiation with respect to $r$ and the Lagrange multiplier $n$
is given by
\begin{equation}\label{B}
n(r)=a(r)N(r)+B^2(r).
\end{equation}
In metric (\ref{A}) the function $N(r)$ plays the role of a lapse
function with respect to the $r$-slicing in the ADM terminology, see
\cite{3} and \cite{Babak}. On the other hand, according to the
relation (\ref{B}), the functions $N$ and $B$ are related to the
Lagrange multiplier $n$ which means that we can arbitrarily choose
them. This is a reflection of this fact that we have freedom in the
definition of the coordinates $r$ and $t$ in the metric (\ref{A}).
Hence, the only independent variables that can be determined by the
Einstein field equations are the functions $a(r)$ and $b(r)$. In
order to write the Hamiltonian the momenta conjugate to these
variables should be evaluated, that is

\begin{equation}\label{C1}
p_a=\frac{\partial L}{\partial
a'}=\frac{2bb'}{\sqrt{n}},\hspace{0.5cm}p_b=\frac{\partial
L}{\partial b'}=2\frac{(2ab'+a'b)}{\sqrt{n}}.\end{equation} Also,
the momentum associated to $n$ vanishes which gives the primary
constraint

\begin{equation}\label{C2}
p_n=\frac{\partial L}{\partial n'}=0.\end{equation} In terms of
these conjugate momenta the canonical Hamiltonian is given by its
standard definition $H=\sum_{q=a,b,n} q'p_q -L$, leading to

\begin{equation}\label{D}
H=\sqrt{n}\left(\frac{p_ap_b}{2b}-\frac{a}{2b^2}p_a^2-2\right)+\Lambda
p_n,
\end{equation}in which due to existence of the constraint
(\ref{C2}), we have added the last term that is the primary
constraints multiplied by an arbitrary functions $\Lambda(r)$. The
Hamiltonian equation for $n$ then reads

\begin{equation}\label{E}
n'=\{n,H\}=\Lambda.
\end{equation}Now, let us restrict ourselves to a certain class of gauges, namely $n
=\mbox{const.}$, which is equivalent to the choice $\Lambda=0$. With
a constant $n$ we assume $n=1$ without losing general character of
the solutions. By this choice, the Hamiltonian equations of motion
for the other variables are as

\begin{eqnarray}\label{F}
\left\{
\begin{array}{ll}
a'=\{a,H\}=\frac{p_b}{2b}-\frac{a}{b^2}p_a,\\\\
p'_a=\{p_a,H\}=\frac{p_a^2}{2b^2},\\\\
b'=\{b,H\}=\frac{p_a}{2b},\\\\
p'_b=\{p_b,H\}=\frac{p_ap_b}{2b^2}-\frac{a}{b^3}p_a^2.
\end{array}
\right.
\end{eqnarray}From the second and third equations of (\ref{F}) we
obtain
\begin{equation}\label{G}
p_a=k_1b,
\end{equation}from which one gets
\begin{equation}\label{H}
b(r)=\frac{k_1}{2}r+k_2,\hspace{5mm}p_a(r)=\frac{k_1^2}{2}r+k_1k_2,
\end{equation}where $k_1$ and $k_2$ are integration constants. Upon
substitution these results into the first and fourth equations of
(\ref{F}) we arrive at the system
\begin{eqnarray}\label{I}
\left\{
\begin{array}{ll}
a'=\frac{1}{k_1r+2k_2}p_b-\frac{2k_1}{k_1r+2k_2}a,\\\\
p'_b=\frac{k_1}{k_1r+2k_2}p_b-\frac{2k_1^2}{k_1r+2k_2}a,
\end{array}
\right.
\end{eqnarray}which results $p'_b=k_1a'$ and then $p_b=k_1a+k_3$.
Therefore,
\begin{equation}\label{J}
a'(r)=-\frac{k_1a+k_3}{k_1r+2k_2},
\end{equation}which after integration we obtain
\begin{equation}\label{K}
a(r)=\frac{k_3}{k_1}+\frac{k_1k_4}{k_1r+2k_2},
\end{equation}and
\begin{equation}\label{L}
p_b(r)=2k_3+\frac{k_1^2k_4}{k_1r+2k_2},
\end{equation}with $k_3$ and $k_4$ being integration constants. Now,
all of the above results should satisfy the constraint equation
$H=0$. Thus, with the help of (\ref{D}) we get $k_1k_3=4$, where we
fix them as $k_1=k_3=2$. Also, $k_2$ and $k_4$ remain arbitrary
where we take their values as $k_2=0$ and $k_4=-2M$ with $M$ being a
constant. Therefore, the metric functions take the form
\begin{equation}\label{M}
a(r)=1-\frac{2M}{r},\hspace{5mm}b(r)=r,
\end{equation}and their conjugate momenta are
\begin{equation}\label{N}
p_a(r)=2r,\hspace{5mm}p_b(r)=4-\frac{4M}{r}.
\end{equation}Finally, with using these relations in (\ref{A}) and (\ref{B}), the
metric is obtained as
\begin{equation}\label{O}
ds^2=-\left(1-\frac{2M}{r}\right)dt^2+N(r)dr^2+2\left[1-\left(1-\frac{2M}{r}\right)N(r)\right]^{1/2}dtdr+r^2\left(d\vartheta^2+\sin^2\vartheta
d\varphi^2\right).
\end{equation}In the final stage we have to eliminate the function $N(r)$. This function should be interpreted as a Lagrange-multiplier and thus,
cannot be considered as a real dynamical variable. As we mentioned
before, one may freely chooses it. From the physical point of view
the function $N(r)$ corresponds to a gauge freedom in choice of
coordinates $r$ and $t$ in the above metric. If we choose the lapse
function as $N(r)=\left(1-\frac{2M}{r}\right)^{-1}$ this metric
takes its canonical form
\begin{equation}\label{P}
ds^2=-\left(1-\frac{2M}{r}\right)dt^2+\left(1-\frac{2M}{r}\right)^{-1}dr^2+r^2\left(d\vartheta^2+\sin^2\vartheta
d\varphi^2\right),
\end{equation}which is nothing but the familiar form for the metric
of the Schwarzschild black hole. However, we may identify the line
element (\ref{O}) with the Eddington-Finkelstein metric
\begin{equation}\label{P1}
ds^2=-\left(1-\frac{2M}{r}\right)dt^2+\frac{4M}{r}dtdr+\left(1+\frac{2M}{r}\right)dr^2+r^2\left(d\vartheta^2+\sin^2\vartheta
d\varphi^2\right),
\end{equation}
for $N(r)=1+2M/r$, or with some other kinds of spherically symmetric
metrics for $N=1$, see \cite{Krous}. In summary, from physical
viewpoint choosing different gauge functions $N(r)$ is actually
looking at a space-time from a different perspective. For example,
the metric (\ref{P1}) can be obtained from (\ref{P}) by introducing
a new time coordinate $\bar{t}=t+2M\ln (r-2M)$ in which the radial
null geodesics (see section 5) become straight lines. In this sense,
the two metrics may differ from some aspects. While the
Schwarzschild metric is singular at $r=2M$ the Eddington-Finkelstein
metric is regular not only at $r=2M$ but also for the whole range
$0<r<2M$. Indeed, the coordinate range is extended from
$2M<r<\infty$ to $0<r<\infty$.

From now on we focus on Schwarzschild black hole metric and to
justify the meaning of the constant $M$, note that the Newtonian
gravitational potential of a point mass $m$ situated at the origin
is given by the relation $\phi=-Gm/r$. On the other hand in the
weak-field limit the $g_{00}$ component of the metric takes the form
$g_{00}=-(1+2\phi/c^2)$ \cite{Inverno}. Therefore, comparing this
with (\ref{P}) we see that: $M=Gm/c^2$. This means that we may
interpret the constant $M$ as due to the mass of the above mentioned
point particle in relativistic units.

\section{Polymerization of the model}
As explained in the second section the method of polymerization is
based on the modification of the Hamiltonian to get a deformed
Hamiltonian $H_{\lambda}$ where $\lambda$ is the deformation
parameter. Quantum polymerization of the spherically symmetric
space-time is studied in \cite{Singh} in which the interior of the
Schwarzschild black hole as described by a Kantowski-Sachs
cosmological model, is quantized by loop quantization method. For
our system this method will be done by applying the transformation
(\ref{PT}) on the Hamiltonian (\ref{D}). However, since all of the
thermodynamical properties of the black hole are encoded in the
function $a(r)$, we will polymerized only the $a(r)$-sector of the
Hamiltonian. So, by means of the transformation

\begin{equation}\label{Q}
a\rightarrow a, \hspace{5mm}p_a\rightarrow
\frac{1}{\lambda}\sin(\lambda p_a),\hspace{5mm}b\rightarrow b
\hspace{5mm}p_b\rightarrow p_b,
\end{equation}
the Hamiltonian takes the form
\begin{equation}\label{R}
H_{\lambda}=-\frac{a}{2b^2}\frac{\sin^2(\lambda
p_a)}{\lambda^2}+\frac{p_b}{2b}\frac{\sin(\lambda p_a)}{\lambda}-2.
\end{equation}As we mentioned earlier, by this one-parameter
$\lambda$-dependent classical theory, we expect to address the
quantum features of the system without a direct reference to the
quantum mechanics. Indeed, here instead of first dealing with the
quantum pictures based on the quantum Hamiltonian operator, one
modifies the classical Hamiltonian according to the transformation
(\ref{PT}) and then deals with classical dynamics of the system with
this deformed Hamiltonian. In the resulting classical system the
discreteness parameter $\lambda$ plays an essential role since its
supports the idea that the $\lambda$-correction to the classical
theory is a signal from quantum gravity. Under these conditions the
Hamiltonian equations of motion for the above Hamiltonian are

\begin{eqnarray}\label{S}
\left\{
\begin{array}{ll}
a'=\{a,H_{\lambda}\}=\frac{p_b}{2b}\cos(\lambda p_a)-\frac{a}{\lambda b^2}\sin(\lambda p_a)\cos(\lambda p_a),\\\\
p'_a=\{p_a,H_{\lambda}\}=\frac{\sin^2(\lambda p_a)}{2\lambda^2b^2},\\\\
b'=\{b,H_{\lambda}\}=\frac{\sin(\lambda p_a)}{2\lambda b},\\\\
p'_b=\{p_b,H_{\lambda}\}=\frac{p_b}{2\lambda b^2}\sin(\lambda
p_a)-\frac{a}{\lambda^2 b^3}\sin^2(\lambda p_a).
\end{array}
\right.
\end{eqnarray}
The second and the third equations of this system give
$\frac{dp_a}{db}=\frac{\sin(\lambda p_a)}{\lambda b}$, integration
of which results: $ b=C_1\tan(\frac{1}{2}\lambda p_a)$, where $C_1$
is an integration constant. We note that in the limit $\lambda
\rightarrow 0$, this relation should back to $b=\frac{1}{2}p_a$,
obtained in the previous section. So, taking this limit fixes the
integration constant as $C_1=\frac{1}{\lambda}$. Therefore,
\begin{equation}\label{T}
b=\frac{1}{\lambda}\tan(\frac{1}{2}\lambda p_a).
\end{equation}Now, we may use this result in the second equation of
(\ref{S}) to arrive at
\begin{equation}\label{U}
p'_a=2\cos^4(\frac{1}{2}\lambda p_a),
\end{equation}whose integral is
\begin{equation}\label{V}
\frac{2}{3}\frac{\tan(\frac{1}{2}\lambda
p_a)}{\lambda}+\frac{1}{3}\frac{\tan(\frac{1}{2}\lambda
p_a)}{\lambda \cos^2(\frac{1}{2}\lambda p_a)}=r.
\end{equation}From (\ref{T}) we get: $\cos^2(\frac{1}{2}\lambda
p_a)=(1+\lambda^2b^2)^{-1}$. With the help of these relations
equation (\ref{V}) takes the following algebraic form for the
function $b(r)$,
\begin{equation}\label{W}
\lambda^2b^3+3b-3r=0,
\end{equation}which admits the exact solution
\begin{equation}\label{X}
b(r)=\frac{\left[3\lambda r+\sqrt{4+9\lambda^2
r^2}\right]^{2/3}-2^{2/3}}{2^{1/3}\lambda \left[3\lambda
r+\sqrt{4+9\lambda^2 r^2}\right]^{1/3}}.
\end{equation}Up to second order of $\lambda$, we have
\begin{equation}\label{Y}
b(r)=r-\frac{1}{3}\lambda^2 r^3+{\cal O}(\lambda^3).
\end{equation}Now, let us back to the first and the fourth equations
of the system (\ref{S}). Using (\ref{T}), they take the form
\begin{equation}\label{Z}
a'=\frac{p_b}{2b}\frac{1-\lambda^2 b^2}{1+\lambda^2
b^2}-\frac{2a}{b}\frac{1-\lambda^2 b^2}{(1+\lambda^2 b^2)^2},
\end{equation}and
\begin{equation}\label{AB}
p'_b=\frac{p_b}{b}\frac{1}{1+\lambda^2
b^2}-\frac{4a}{b}\frac{1}{(1+\lambda^2 b^2)^2},
\end{equation}in which we have used the trigonometric relations:
$\sin(\lambda p_a)=2\lambda b/(1+\lambda^2 b^2)$ and $\cos^2(\lambda
p_a)=(1-\lambda^2 b^2)/(1+\lambda^2 b^2)$. From these two equations
we get
\begin{equation}\label{AC}
p'_b=\frac{2}{1-\lambda^2 b^2}a',
\end{equation}where up to second order of $\lambda$, using (\ref{Y}) is
\begin{equation}\label{AD}
p'_b=2\left[1+\lambda^2 r^2+{\cal O}(\lambda^3)\right]a',
\end{equation}and thus
\begin{equation}\label{AE}
p_b=2\int (1+\lambda^2 r^2)a'dr.
\end{equation}We may use this relation in (\ref{Z}) to get a
differential equation for $a(r)$. However, since the resulting
equation seems to be too complicated to have an exact solution, we
rely on an approximation according to which we ignore all powers of
$\lambda$ in the r.h.s. of (\ref{Z}) and so obtain
\begin{equation}\label{AF}
a'(r)=\frac{1}{r}\int (1+\lambda^2 r^2)a'dr-\frac{2}{r}a,
\end{equation}or after differentiation of both sides
\begin{equation}\label{AG}
ra''(r)=(\lambda^2 r^2-2)a',
\end{equation}with solution
\begin{equation}\label{AH}
a(r)=C_2+C_3\left[-\frac{1}{r}e^{\lambda^2
r^2/2}+\lambda\sqrt{\frac{\pi}{2}}\mbox{erfi}\left(\frac{\lambda
r}{\sqrt{2}}\right)\right],
\end{equation}where $C_2$ and $C_3$ are two integration constant and $\mbox{erfi}(z)$ is the imaginary error function. Up
to second order of $\lambda$ this expression has the form
\begin{equation}\label{AI}
a(r)=C_2-\frac{C_3}{r}+\frac{1}{2}C_3 \lambda^2 r,
\end{equation}comparison of which with (\ref{M}) suggests that the
integration constants should fix as $C_2=1$ and $C_3=2M$. So,
\begin{equation}\label{AJ}
a(r)=1+2M\left[-\frac{1}{r}e^{\lambda^2
r^2/2}+\lambda\sqrt{\frac{\pi}{2}}\mbox{erfi}\left(\frac{\lambda
r}{\sqrt{2}}\right)\right].
\end{equation}Therefore, by choosing the lapse function in the form
$N(r)=a^{-1}(r)$, (see the discussion after equation (\ref{O})), the
polymerized metric takes the form

\begin{equation}\label{AK}
ds^2=-a(r)dt^2+a^{-1}(r)dr^2+b^2(r)\left(d\vartheta^2+\sin^2\vartheta
d\varphi^2\right),
\end{equation}where $a(r)$ and $b(r)$ are given in (\ref{AJ}) and
(\ref{Y}) respectively. It is seen that in the limit
$\lambda\rightarrow 0$ the line element (\ref{AK}) returns to the
usual Schwarzschild metric (\ref{P}). However, its asymptotic
behavior which comes from the expansion

\begin{equation}\label{AK1}
a(r)=1-\frac{2M}{r}+M\lambda^2 r\left[1+\frac{1}{12}(\lambda
r)^2+\frac{1}{120}(\lambda r)^4+...\right],
\end{equation}shows that in spite of the Schwarzschild case, the
metric is not flat for large values of $r$. Later in this section,
we attribute such a asymptotic behavior to the matter field that
created this metric. In what follows, we will deal with the physical
properties, including thermodynamics, of the space-time (\ref{AK}).
Since such properties of a black hole can be derived from its
geometry, we expect that the deformed form of these properties
return to their ordinary form in the limit $\lambda\rightarrow 0$.

At first, let us take a look at the horizon(s) radius of the metric
(\ref{AK}) which may be deduced from the roots of equation $a(r)=0$.
Up to the leading order of parameter $\lambda$, the positive root of
this equation is

\begin{equation}\label{AL}
r_H\simeq\frac{\sqrt{1+8M^2\lambda^2}-1}{2M\lambda^2}.
\end{equation}On the other hand since $\frac{da(r)}{dr}=\frac{2Me^{\lambda^2 r^2/2}}{r^2}>0$, the function $a(r)$ is monotonically
increasing and thus the metric can not have more than one horizon
whose radius is approximately given in equation (\ref{AL}). To see
the behavior of the above metric near the Schwarzschild essential
singularity $r=0$, we may evaluate some scalars associated to the
metric such as Ricci scalar $R$, $R_{\mu \nu}R^{\mu \nu}$ and the
Kretschmann scalar $K=R_{\mu \nu \sigma \delta}R^{\mu \nu \sigma
\delta}$. A straightforward calculation shows that

\begin{equation}\label{AM}
R=\frac{2\lambda M}{r^2}e^{\frac{\lambda^2 r^2}{2}}\left[\lambda
r+2\sqrt{2}F\left(\frac{\lambda r}{\sqrt{2}}\right)\right],
\end{equation}where $F(x)=e^{-x^2}\int_0^x e^{y^2}dy$, is the Dawson
function. Near $r=0$, the above relation behaves as $6M\lambda^2/r$,
so, given that the value of the parameter $\lambda$ is also very
small we have $\lim_{r\rightarrow 0} R \simeq {\cal O}(\lambda)$.
Computing of the scalar $R_{\mu \nu}R^{\mu \nu}$ shows the similar
behavior near $r=0$, while the Kretschmann scalar takes the form

\begin{equation}\label{AN}
K=\frac{4 M^2 e^{\lambda ^2 r^2}}{r^6} \left[4 \lambda ^2 r^2
\left(2 F\left(\frac{r \lambda }{\sqrt{2}}\right)^2-1\right)-8
\sqrt{2} \lambda  r F\left(\frac{r \lambda
}{\sqrt{2}}\right)+\lambda ^4 r^4+12\right],
\end{equation}
which behaves as $K\simeq
\frac{48M^2}{r^6}+\frac{20M^2\lambda^4}{3r^2}+{\cal O}(\lambda^5)$.
Thus near $r=0$ we have $K\simeq \frac{48M^2}{r^6}$. This shows that
the space-time described by the metric (\ref{AK}) has an essential
singularity at $r=0$, which can not be removed by a coordinate
transformation.

Now, let us investigate the properties of the matter corresponds to
the metric (\ref{AK}). Considering the Einstein equations
$G^{\mu}_{\nu}=R^{\mu}_{\nu}-\frac{1}{2}R\delta^{\mu}_{\nu}\sim
T^{\mu}_{\nu}$, the components of the energy-momentum tensor become

\begin{eqnarray}\label{AO}
T^{\mu}_{\nu}&=&\mbox{diag}\left(-\rho,p_r,p_{\bot},p_{\bot}\right)\\
\nonumber &=& \mbox{diag}\left(-\frac{\sqrt{2 \pi } \lambda  M
\text{erfi}\left(\frac{\lambda
r}{\sqrt{2}}\right)}{r^2},-\frac{\sqrt{2 \pi } \lambda  M
\text{erfi}\left(\frac{\lambda
r}{\sqrt{2}}\right)}{r^2},-\frac{\lambda ^2 M e^{\frac{\lambda ^2
r^2}{2}}}{r},-\frac{\lambda ^2 M e^{\frac{\lambda ^2
r^2}{2}}}{r}\right).
\end{eqnarray}
Before going any further, a remark is in order. The usual
Schwarzschild metric is often considered a vacuum solution since it
solves $R_{\mu \nu}=0$ which is equivalent to the Einstein vacuum
field equations $G_{\mu \nu}=0$. However, as (\ref{AO}) explicitly
shows the polymer corrected metric (\ref{AK}) is not a vacuum
solution. Then, a question arises: What mechanism made it possible
starting from a vacuum solution we get a non-vacuum solution? To
deal with this question note that any vacuum solution must be found
in the absence of matter, strictly speaking, only the Minkowski
metric can be considered as a vacuum solution. As shown in
\cite{Balasin}, in the Schwarzschild case there is a source term
(energy-momentum tensor) concentrated on the origin, the origin
which usually excluded from the space-time manifold. So we are faced
with an unacceptable physical situation in which a curved metric is
generated by a zero energy-momentum tensor. In \cite{Balasin} with
more accurate calculations based on distributional techniques the
energy-momentum tensor of the Schwarzschild geometry is obtained and
it has been shown that its Ricci scalar is equal to $8\pi M
\delta(r)$ which yields an energy-momentum tensor proportional to
$M\delta(r)$. Now, what is happening in the effective theories such
as noncommutative, see \cite{Nic}, and polymeric counterparts of the
Schwarzschild solution is that the concentrated matter on the origin
will spread throughout space by the polymer parameter $\lambda$ (or
noncommutative parameter $\theta$ in noncommutative theories).

The energy-momentum tensor (\ref{AO}) shows a fluid with radial
pressure $p_r=-\rho$ and tangential pressure
$p_{\bot}=-\rho-\frac{r}{2}\partial_r \rho$. In comparison with the
conventional perfect fluid with isotropic pressure the above
energy-momentum tensor shows an unusual behavior since its pressure
exhibits an anisotropic behavior. At short distances the difference
between $p_r$ and $p_{\bot}$ is of order $\lambda^2$, which shows
the fluid behaves approximately like a perfect fluid. However, when
$r$ grows the anisotropy between the pressure's components increases
and the behavior of the fluid is far from the perfect fluid
behavior. The non-vanishing radial pressure of the above anisotropic
fluid may be interpreted as a result of the quantum fluctuation of
given space-time. The large amount of this pressure near the origin
prevents the matter collapsing into the this point. Such an unusual
equation of state for fluids is also appeared in the noncommutative
theories of black holes \cite{Nic}. In view of the validity of the
energy conditions, we see that

\begin{equation}\label{AP}
\rho+p_r+2p_{\bot}=-2\frac{\lambda ^2 M e^{\frac{\lambda ^2
r^2}{2}}}{r}<0,
\end{equation}which shows the violation of the strong energy
condition for this exotic distribution of matter. On the other hand,
in view of the weak energy conditions, while the relation
$\rho+p_r\geq 0$ is always satisfied, the condition
$\rho+p_{\bot}\geq 0$ is violated for $r
> {\cal O}(1/\lambda)$. The violation of the energy conditions shows
that the classical description of this type of matter field is not
credible and thus the corresponding gravity should be described by
an effective quantum theory (here the polymerized theory) rather
than the usual general relativity.

\begin{figure}
\includegraphics[width=2.5in]{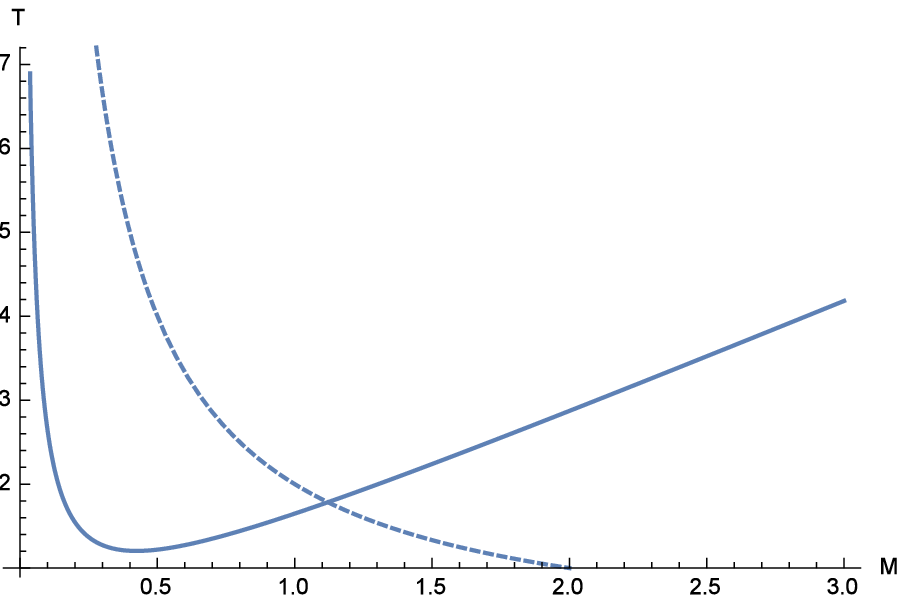}\hspace{3cm}\includegraphics[width=2.5in]{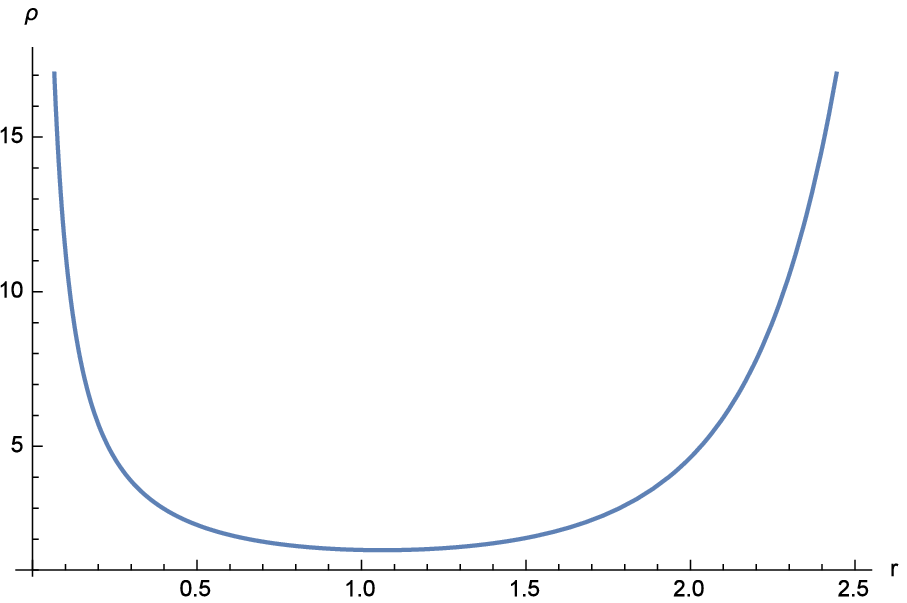}
\caption{Left: Temperature versus mass. The solid line shows the
qualitative behavior of the relation (\ref{AR}) while the dashed
line refers to the conventional temperature of the Schwarzschild
black hole. Right: The density of the matter distribution versus
$r$. The figures are plotted for $\lambda=0.1$ and
$M=1$.}\label{fig1}
\end{figure}

\begin{figure}
\includegraphics[width=2.5in]{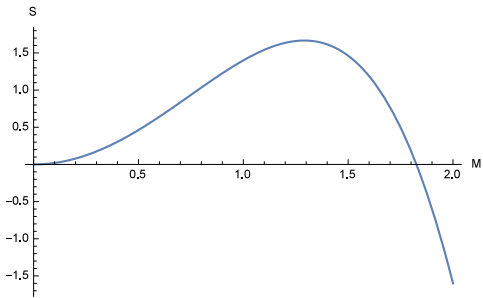}
\caption{The entropy versus mass. The figure is plotted for
$\lambda=0.1$ and $M=1$.}\label{fig2}
\end{figure}

Finally, let us take a quick look at thermodynamics of the metric
(\ref{AK}). According to the Hawking formulation the black hole's
temperature is proportional to the surface gravity at the black hole
horizon. It can be shown that for a diagonal metric such as
(\ref{AK}) the surface gravity is \cite{Hervik}

\begin{equation}\label{AQ}
\kappa=\sqrt{-\frac{1}{4}g^{tt}g^{rr}\left(\frac{\partial
g_{tt}}{\partial r}\right)^2}=\frac{M}{r^2}e^{\frac{\lambda^2
r^2}{2}}.
\end{equation}Evaluating this expression at the horizon radius
(\ref{AL}) gives the temperature as

\begin{equation}\label{AR}
T\propto \frac{4 \lambda ^4 M^3 e^{\frac{\left(\sqrt{8 \lambda ^2
M^2+1}-1\right)^2}{8 \lambda ^2 M^2}}}{\left(\sqrt{8 \lambda ^2
M^2+1}-1\right)^2}.
\end{equation}In figure \ref{fig1} we have plotted the qualitative
behavior of the above results. As this figure shows near the origin
the matter has a dense core like its conventional Schwarzschild
counterpart. Thus, the temperature changes like Schwarzschild in
this regime. However, in a global look, the exotic properties of the
matter cause different behavior for temperature. Unlike the usual
Schwarzschild case, by decreasing the mass, the radiation
temperature first decreases to a minimum value and then exhibits the
normal behavior, i.e. the temperature increase while the mass is
decreasing. The reason for this abnormal behavior in the temperature
of the radiation may be found in the nature of the dark energy-like
of the matter field described by the energy-momentum tensor
(\ref{AO}). Now, by the second law of thermodynamics $dS=dM/T$, we
may compute the entropy as

\begin{equation}\label{AS}
S=\int \frac{e^{-\frac{\left(\sqrt{8 \lambda ^2 M^2+1}-1\right)^2}{8
\lambda ^2 M^2}} \left(\sqrt{8 \lambda ^2 M^2+1}-1\right)^2}{4
\lambda ^4 M^3} dM.
\end{equation}We see that this integral cannot be evaluated analytically. In figure \ref{fig2}, employing numerical methods, we have shown the approximate
behavior of the entropy for typical values of the parameters. As the
figure shows with decreasing mass, the entropy grows from negative
values up to a maximum positive value and then behaves like the
Schwarzschild case and decreases to zero.

\section{Geodesics of the polymerized metric}

\begin{figure}
\includegraphics[width=2.5in]{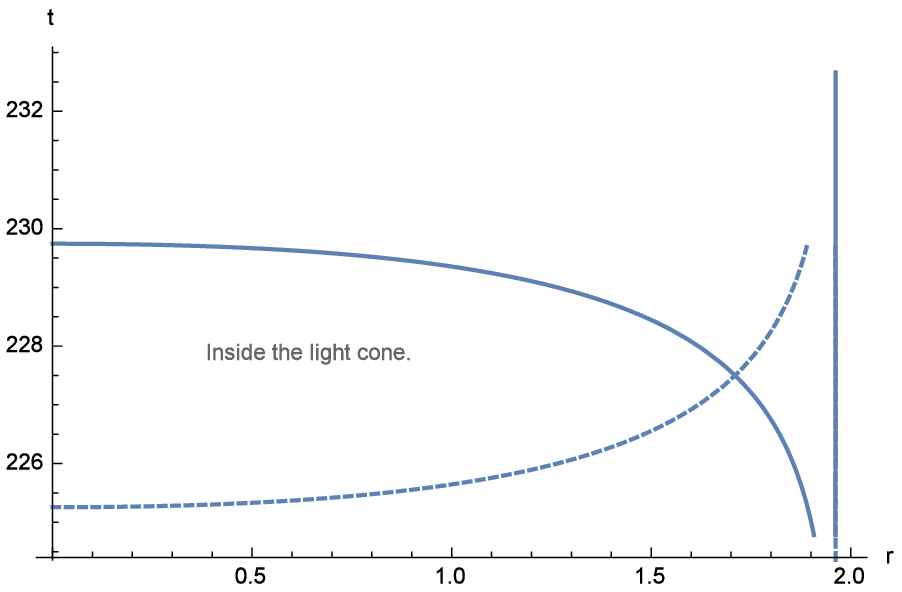}\includegraphics[width=2.5in]{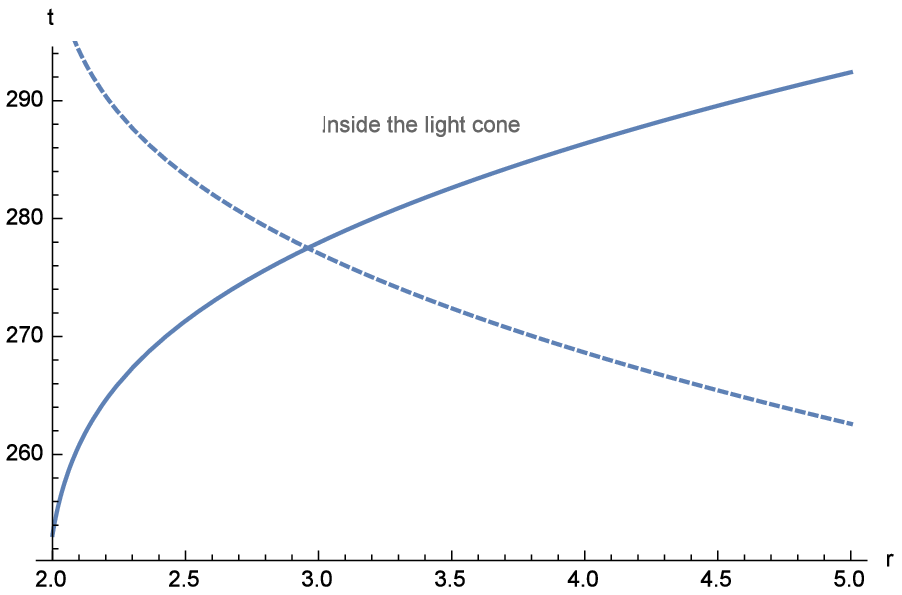}
\caption{Left: The outgoing (dashed line) and incoming (solid line)
geodesics for $r<r_H$. Right: The outgoing (solid line) and incoming
(dashed line) geodesics $r>r_H$. The figures are plotted for
$\lambda=0.1$ and $M=1$.}\label{fig3}
\end{figure}

In this section we are going to study how light and particles will
move in the geometrical background given by metric (\ref{AK}). This
is important because from the classical trajectories of light or
falling particles we understand that the corresponding space-time
behaves really like a black hole. First, consider the radial null
geodesics defined by $ds=0$ and $d\vartheta=d\varphi=0$. Therefore,
we have

\begin{equation}\label{AT}
-a(r)dt^2+a^{-1}(r)dr^2=0 \Rightarrow dt=\pm \frac{dr}{a(r)}.
\end{equation}In order to get an analytical solution we use the
approximate $a(r)\sim 1-\frac{2M}{r}+M\lambda^2 r$, for which we
obtain

\begin{equation}\label{AU}
t-t_0=\pm\frac{\frac{2|\tanh ^{-1}\left(\frac{2 M r \lambda
^2+1}{\sqrt{8 M^2 \lambda ^2+1}}\right)|}{\sqrt{8 \lambda ^2
M^2+1}}+\log |M r^2 \lambda ^2-2 M+r|}{2 \lambda ^2 M}.
\end{equation}For $r>r_H$, the above expression with positive (negative) sign
shows that $r$ increases (decreases) as $t$ increases and thus the
corresponding curve is an {\it outgoing} ({\it incoming}) radial
null geodesics. For $r<r_H$, the situation is reversed, i.e., the
positive and negative signs correspond to the incoming and outgoing
curves respectively, see figure \ref{fig3}. A glance at this figure
makes it clear that none of the null geodesics can pass through the
horizon which shows the black hole nature of the underlying
space-time. It is clear that in comparison with region $r>r_H$ the
local light cones tip over in region $r<r_H$. This is because that
while the coordinates $r$ and $t$ are space-like and time-like
respectively in region $r>r_H$, in region $r<r_H$ they reverse their
character. The orientation of the light cones inside the horizon
shows that nothing can stay at rest in this region but will be
forced to move towards the black hole center.

\begin{figure}
\includegraphics[width=2.5in]{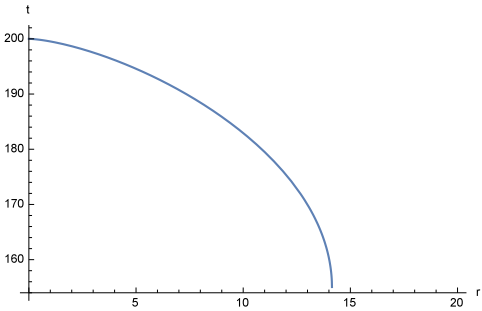}
\caption{The trajectory of an infalling particle in terms of the
proper time $\tau$. The particle falls continuously to the
singularity $r=0$ in a finite proper time. The figure is plotted for
$\lambda=0.1$ and $M=1$.}\label{fig4}
\end{figure}

To complete our geodesics analysis, let us now consider the radial
trajectory of a falling free particle. It moves along the time-like
geodesics which results the following equations of motion
\cite{Inverno}

\begin{equation}\label{AV}
a(r)\dot{t}=k,
\end{equation}
\begin{equation}\label{AW}
a(r)\dot{t}^2-a^{-1}(r)\dot{r}^2=1,
\end{equation}where a dot denotes differentiation with respect to
the proper time $\tau$ and $k$ is a constant that depends on the
initial conditions. If we assume that the particle begins to fall
with zero initial velocity from a distance $r_0$ for which
$a(r_0)=1$, then $k=1$. Also, for the motion around this region we
have $\dot{t}\simeq 1\Rightarrow t\simeq \tau$. Therefore, we may
analysis the path of the particle in view a co-moving observer which
uses the proper time. Then, equations (\ref{AV}) and (\ref{AW}) give
\begin{equation}\label{AX}
\frac{d\tau}{dr}=-\left(\frac{r}{2M-M\lambda^2r^2}\right)^{1/2},
\end{equation}in which we have used the same previous approximation for
$a(r)$. Upon integration we get

\begin{equation}\label{AY}
\tau-\tau_0=-\frac{1}{3} r \sqrt{4-2 \lambda ^2 r^2} \,
_2F_1\left(\frac{1}{2},\frac{3}{4};\frac{7}{4};\frac{r^2 \lambda
^2}{2}\right) \sqrt{\frac{r}{2 M-\lambda ^2 M r^2}},
\end{equation}where $\tau_0$ is an integration constant and $\,
_2F_1\left(a,b;c;z\right)$ is a Hypergeometric function. The above
equation shows that in view of the proper observer no singular
behavior occurs at the horizon radius and the particle falls to the
center of the black hole, see figure \ref{fig4}. If instead one
describes the motion in terms of the coordinate time $t$, the
situation becomes like the null geodesics, i.e., in view of a
distance observer the particle can not pass through the horizon and
it takes infinite time for the falling particle to reach the
horizon, so that the $r_H$ is approached but never passed. All of
these results show that in terms of light and  particles motion the
space-time given by the metric (\ref{AK}) behaves like a black hole
as its Schwarzschild counterpart.

\section{Summary}In this paper we have studied the classical polymerization procedure applied on the Schwarzschild
metric. This procedure is based on a classical transformation under
which the momenta are transformed like their polymer quantum
mechanical counter part. After a brief review of the polymer
representation of quantum mechanics, we have introduced the
classical polymerization by means of which the Hamiltonian of the
theory under consideration gets modification in such a way that a
parameter $\lambda$, coming from polymer quantization, plays the
role of a deformation parameter. In order to apply this mechanism on
the Schwarzschild black hole, we first presented a Hamiltonian
function for a general spherically symmetric space-time and showed
that the resulting Hamiltonian equations yield the conventional
Schwarzschild metric. Then, we have applied the polymerization on
this minisuperspace model and solved the Hamiltonian equations once
again to achieve the polymer corrected Schwarzschild metric. We saw
that while the usual Schwarzschild metric is a vacuum spherical
symmetric solution of the Einstein equations, this is not the case
for its polymerized version obtained by the above mentioned method.
Interestingly, the energy-momentum tensor of the matter field
corresponding to the polymerized metric has anisotropic negative
pressure sector with a dark energy-like equation of state. As
expected, the unusual behavior of such a matter field resulted an
uncommon behavior for the thermodynamical quantities like
temperature and entropy in comparison with the traditional
Schwarzschild solution. Finally, to clarify that the polymerized
metric has also the black hole nature, we have investigated the null
geodesics and verified that the outgoing and incoming geodesics
curves can never pass through the horizon. Also, we proved that in
view of a co-moving observer which uses the proper time, an
infalling particle continuously falls to the center $r=0$, without
experiencing something passing through the horizon. All these
indicate that the underlying space-time in which the light and
particles are traveling is really a black hole.

\vskip 0.2 in \textit{Acknowledgments:} This work has been supported
financially by Research Institute for Astronomy and Astrophysics of
Maragha (RIAAM) under research project NO. 1/5237-107.

\end{document}